\documentstyle[12pt,a4,epsf,epsfig,amsmath,psfrag]{article}
\font\ninerm=cmr9

\newcommand{\ee}{\epem}

\newcommand{\WW}{\WpWm}

\def\be{\begin{equation}}
\def\ee{\end{equation}}
\def\bea{\begin{eqnarray}}
\def\eea{\end{eqnarray}}
\def\beas{\begin{eqnarray*}}
\def\eeas{\end{eqnarray*}}

\def\a_la_ligne{}

\def\W{\mbox{$\hbox{\rm W}$}}

\def\WpWm{\mbox{$\rm \W^+\W^-$}}

\def\goes{\mbox{$\rightarrow$}}

\def\epem{\mbox{$\mathrm{e^+ e^-}$}}

\def\gg{\mbox{gg}}

\def\ccbar{\mbox{${\mathrm {c\bar{c}}}$}}

\def\bbbar{\mbox{${\mathrm {b\bar{b}}}$}}

\def\mH{\mbox{$m_{\mathrm H}$}}

\def\mW{\mbox{$m_{\mathrm W}$}}

\def\EPJ#1#2#3{{ Eur.\ Phys.\ J. }{\bf #1 }(#2) #3}

\def\HZHA{{\sc HZHA}}

\def\2HDM{2HDM} 

\def\4jets{Four jets final state}

\def\S_dens{${\cal S}_{m_{\rm H}}$}
\def\B_dens{${\cal B}_{m_{\rm H}}$}

\def\m34{$\rm m_{34}$}

\def\mw{\mW}

\def\Ejmin{$E_{\rm {jet}}^{\rm {min}}$}
\def\Ejmax{$E_{\rm {jet}}^{\rm {max}}$}
\def\Ejminthe{$E_{\rm{jet}}^{\rm{min}} \theta_{ij}$}
\def\EllWi{$\rm {\cal E}_{WW}$}
\def\EllhZi{$\rm {\cal E}_{HZ}$}

\def\Y34{$\rm y_{34}$}

\def\mij{$m_{ij}$}
\def\mkl{$m_{kl}$}
\def\minijkl{$\underset{i,j,k,l=1,4}{{\min}}$}
\def\N95{$\rm \overline{N}_{95}$} 

\def\xihad{$\xi_{\rm {had}}^2$}
\def\xitau{$\xi^{2}_\tau$}
\def\xihadplane{\hbox{(\mH, \xihad)}}
\def\xitauplane{\hbox{(\mH, \xitau)}}
\def\mc2{\,GeV/$c^2$}
\def\mevc2{\,MeV/$c^2$}
\def\SM{Standard Model}
\def\Hhad{$\rm{H} \to {\rm hadrons}$}

\def\sm1{${\cal S}_{m_1}$}  
\def\sm2{${\cal S}_{m_2}$}  
 
\def\bm1{${\cal B}_{m_1}$}  
\def\bm2{${\cal B}_{m_2}$}  
\begin{document}
%
%
\thispagestyle{empty}

\date{}
\title{ \null\vspace{1cm} \bf A flavour-independent Higgs boson
      search\\ in  ${\boldsymbol{ {\rm e}^+{\rm e}^-}}$  collisions at 
${\boldsymbol{\sqrt{s}}}$ up to
      209\,GeV
\vspace{1cm}}
\author{The ALEPH Collaboration$^*)$}

\thispagestyle{empty}
\maketitle
\thispagestyle{empty}
\begin{picture}(160,1)
\put(0,115){\rm ORGANISATION EUROP\'EENNE POUR LA RECHERCHE NUCL\'EAIRE
(CERN)}
\put(30,110){\rm Laboratoire Europ\'een pour la Physique des Particules}
\put(125,94){\parbox[t]{45mm}{\tt CERN-EP/2002-027}}
\put(125,88){\parbox[t]{45mm}{\tt 25-Apr-2002}}
\end{picture}

\vspace{.2cm}
\begin{abstract}
A search for the Higgsstrahlung process ${\rm e}^+{\rm e}^- \to {\rm HZ}$
 is carried out, covering decays of the Higgs boson into any quark pair,
  a gluon pair or a tau pair.
  The analysis is based on the $630\,{\rm pb}^{-1}$ of data collected by 
the ALEPH detector at LEP at  centre-of-mass energies from 189 to 209\,GeV.
 A 95\% C.L. lower mass limit of 109.1\mc2 is obtained 
 for a Higgs boson cross section equal to that expected from  the \SM{} if 
the Higgs boson decays exclusively into hadrons and/or taus, 
irrespective of the relative branching fractions.
\vspace{.2cm}
\end{abstract}

\vfill
\centerline{\it Submitted to Physics Letters B}
\vskip .5cm
\noindent
--------------------------------------------\hfill\break
{\ninerm $^*)$ \footnotesize See next pages for the list of authors}

\pagestyle{empty}

\eject

\pagestyle{empty}
\newpage
\small
%
%
\newlength{\saveparskip}
\newlength{\savetextheight}
\newlength{\savetopmargin}
\newlength{\savetextwidth}
\newlength{\saveoddsidemargin}
\newlength{\savetopsep}
\setlength{\saveparskip}{\parskip}
\setlength{\savetextheight}{\textheight}
\setlength{\savetopmargin}{\topmargin}
\setlength{\savetextwidth}{\textwidth}
\setlength{\saveoddsidemargin}{\oddsidemargin}
\setlength{\savetopsep}{\topsep}
%
%
\setlength{\parskip}{0.0cm}
\setlength{\textheight}{25.0cm}
\setlength{\topmargin}{-1.5cm}
\setlength{\textwidth}{16 cm}
\setlength{\oddsidemargin}{-0.0cm}
\setlength{\topsep}{1mm}
\pretolerance=10000
\centerline{\large\bf The ALEPH Collaboration}
\footnotesize
\vspace{0.5cm}
{\raggedbottom
\begin{sloppypar}
\samepage\noindent
A.~Heister,
S.~Schael
\nopagebreak
\begin{center}
\parbox{15.5cm}{\sl\samepage
Physikalisches Institut das RWTH-Aachen, D-52056 Aachen, Germany}
\end{center}\end{sloppypar}
\vspace{2mm}
\begin{sloppypar}
\noindent
R.~Barate,
R.~Bruneli\`ere,
I.~De~Bonis,
D.~Decamp,
C.~Goy,
S.~Jezequel,
J.-P.~Lees,
F.~Martin,
E.~Merle,
\mbox{M.-N.~Minard},
B.~Pietrzyk,
B.~Trocm\'e
\nopagebreak
\begin{center}
\parbox{15.5cm}{\sl\samepage
Laboratoire de Physique des Particules (LAPP), IN$^{2}$P$^{3}$-CNRS,
F-74019 Annecy-le-Vieux Cedex, France}
\end{center}\end{sloppypar}
\vspace{2mm}
\begin{sloppypar}
\noindent
G.~Boix,$^{25}$
S.~Bravo,
M.P.~Casado,
M.~Chmeissani,
J.M.~Crespo,
E.~Fernandez,
M.~Fernandez-Bosman,
Ll.~Garrido,$^{15}$
E.~Graug\'{e}s,
J.~Lopez,
M.~Martinez,
G.~Merino,
R.~Miquel,$^{4}$
Ll.M.~Mir,$^{4}$
A.~Pacheco,
D.~Paneque,
H.~Ruiz
\nopagebreak
\begin{center}
\parbox{15.5cm}{\sl\samepage
Institut de F\'{i}sica d'Altes Energies, Universitat Aut\`{o}noma
de Barcelona, E-08193 Bellaterra (Barcelona), Spain$^{7}$}
\end{center}\end{sloppypar}
\vspace{2mm}
\begin{sloppypar}
\noindent
A.~Colaleo,
D.~Creanza,
N.~De~Filippis,
M.~de~Palma,
G.~Iaselli,
G.~Maggi,
M.~Maggi,
S.~Nuzzo,
A.~Ranieri,
G.~Raso,$^{24}$
F.~Ruggieri,
G.~Selvaggi,
L.~Silvestris,
P.~Tempesta,
A.~Tricomi,$^{3}$
G.~Zito
\nopagebreak
\begin{center}
\parbox{15.5cm}{\sl\samepage
Dipartimento di Fisica, INFN Sezione di Bari, I-70126 Bari, Italy}
\end{center}\end{sloppypar}
\vspace{2mm}
\begin{sloppypar}
\noindent
X.~Huang,
J.~Lin,
Q. Ouyang,
T.~Wang,
Y.~Xie,
R.~Xu,
S.~Xue,
J.~Zhang,
L.~Zhang,
W.~Zhao
\nopagebreak
\begin{center}
\parbox{15.5cm}{\sl\samepage
Institute of High Energy Physics, Academia Sinica, Beijing, The People's
Republic of China$^{8}$}
\end{center}\end{sloppypar}
\vspace{2mm}
\begin{sloppypar}
\noindent
D.~Abbaneo,
P.~Azzurri,
T.~Barklow,$^{30}$
O.~Buchm\"uller,$^{30}$
M.~Cattaneo,
F.~Cerutti,
B.~Clerbaux,$^{34}$
H.~Drevermann,
R.W.~Forty,
M.~Frank,
F.~Gianotti,
T.C.~Greening,$^{26}$
J.B.~Hansen,
J.~Harvey,
D.E.~Hutchcroft,
P.~Janot,
B.~Jost,
M.~Kado,$^{2}$
P.~Mato,
A.~Moutoussi,
F.~Ranjard,
L.~Rolandi,
D.~Schlatter,
G.~Sguazzoni,
W.~Tejessy,
F.~Teubert,
A.~Valassi,
I.~Videau,
J.J.~Ward
\nopagebreak
\begin{center}
\parbox{15.5cm}{\sl\samepage
European Laboratory for Particle Physics (CERN), CH-1211 Geneva 23,
Switzerland}
\end{center}\end{sloppypar}
\vspace{2mm}
\begin{sloppypar}
\noindent
F.~Badaud,
S.~Dessagne,
A.~Falvard,$^{20}$
D.~Fayolle,
P.~Gay,
J.~Jousset,
B.~Michel,
S.~Monteil,
D.~Pallin,
J.M.~Pascolo,
P.~Perret
\nopagebreak
\begin{center}
\parbox{15.5cm}{\sl\samepage
Laboratoire de Physique Corpusculaire, Universit\'e Blaise Pascal,
IN$^{2}$P$^{3}$-CNRS, Clermont-Ferrand, F-63177 Aubi\`{e}re, France}
\end{center}\end{sloppypar}
\vspace{2mm}
\begin{sloppypar}
\noindent
J.D.~Hansen,
J.R.~Hansen,
P.H.~Hansen,
B.S.~Nilsson
\nopagebreak
\begin{center}
\parbox{15.5cm}{\sl\samepage
Niels Bohr Institute, 2100 Copenhagen, DK-Denmark$^{9}$}
\end{center}\end{sloppypar}
\vspace{2mm}
\begin{sloppypar}
\noindent
A.~Kyriakis,
C.~Markou,
E.~Simopoulou,
A.~Vayaki,
K.~Zachariadou
\nopagebreak
\begin{center}
\parbox{15.5cm}{\sl\samepage
Nuclear Research Center Demokritos (NRCD), GR-15310 Attiki, Greece}
\end{center}\end{sloppypar}
\vspace{2mm}
\begin{sloppypar}
\noindent
A.~Blondel,$^{12}$
\mbox{J.-C.~Brient},
F.~Machefert,
A.~Roug\'{e},
M.~Swynghedauw,
R.~Tanaka
\linebreak
H.~Videau
\nopagebreak
\begin{center}
\parbox{15.5cm}{\sl\samepage
Laboratoire de Physique Nucl\'eaire et des Hautes Energies, Ecole
Polytechnique, IN$^{2}$P$^{3}$-CNRS, \mbox{F-91128} Palaiseau Cedex, France}
\end{center}\end{sloppypar}
\vspace{2mm}
\begin{sloppypar}
\noindent
V.~Ciulli,
E.~Focardi,
G.~Parrini
\nopagebreak
\begin{center}
\parbox{15.5cm}{\sl\samepage
Dipartimento di Fisica, Universit\`a di Firenze, INFN Sezione di Firenze,
I-50125 Firenze, Italy}
\end{center}\end{sloppypar}
\vspace{2mm}
\begin{sloppypar}
\noindent
A.~Antonelli,
M.~Antonelli,
G.~Bencivenni,
F.~Bossi,
G.~Capon,
V.~Chiarella,
P.~Laurelli,
G.~Mannocchi,$^{5}$
G.P.~Murtas,
L.~Passalacqua
\nopagebreak
\begin{center}
\parbox{15.5cm}{\sl\samepage
Laboratori Nazionali dell'INFN (LNF-INFN), I-00044 Frascati, Italy}
\end{center}\end{sloppypar}
\vspace{2mm}
\begin{sloppypar}
\noindent
J.~Kennedy,
J.G.~Lynch,
P.~Negus,
V.~O'Shea,
A.S.~Thompson
\nopagebreak
\begin{center}
\parbox{15.5cm}{\sl\samepage
Department of Physics and Astronomy, University of Glasgow, Glasgow G12
8QQ,United Kingdom$^{10}$}
\end{center}\end{sloppypar}
\vspace{2mm}
\pagebreak
\begin{sloppypar}
\noindent
S.~Wasserbaech
\nopagebreak
\begin{center}
\parbox{15.5cm}{\sl\samepage
Department of Physics, Haverford College, Haverford, PA 19041-1392, U.S.A.}
\end{center}\end{sloppypar}
\vspace{2mm}
\begin{sloppypar}
\noindent
R.~Cavanaugh,$^{33}$
S.~Dhamotharan,$^{21}$
C.~Geweniger,
P.~Hanke,
V.~Hepp,
E.E.~Kluge,
G.~Leibenguth,
A.~Putzer,
H.~Stenzel,
K.~Tittel,
M.~Wunsch$^{19}$
\nopagebreak
\begin{center}
\parbox{15.5cm}{\sl\samepage
Kirchhoff-Institut f\"ur Physik, Universit\"at Heidelberg, D-69120
Heidelberg, Germany$^{16}$}
\end{center}\end{sloppypar}
\vspace{2mm}
\begin{sloppypar}
\noindent
R.~Beuselinck,
W.~Cameron,
G.~Davies,
P.J.~Dornan,
M.~Girone,$^{1}$
R.D.~Hill,
N.~Marinelli,
J.~Nowell,
S.A.~Rutherford,
J.K.~Sedgbeer,
J.C.~Thompson,$^{14}$
R.~White
\nopagebreak
\begin{center}
\parbox{15.5cm}{\sl\samepage
Department of Physics, Imperial College, London SW7 2BZ,
United Kingdom$^{10}$}
\end{center}\end{sloppypar}
\vspace{2mm}
\begin{sloppypar}
\noindent
V.M.~Ghete,
P.~Girtler,
E.~Kneringer,
D.~Kuhn,
G.~Rudolph
\nopagebreak
\begin{center}
\parbox{15.5cm}{\sl\samepage
Institut f\"ur Experimentalphysik, Universit\"at Innsbruck, A-6020
Innsbruck, Austria$^{18}$}
\end{center}\end{sloppypar}
\vspace{2mm}
\begin{sloppypar}
\noindent
E.~Bouhova-Thacker,
C.K.~Bowdery,
D.P.~Clarke,
G.~Ellis,
A.J.~Finch,
F.~Foster,
G.~Hughes,
R.W.L.~Jones,
M.R.~Pearson,
N.A.~Robertson,
M.~Smizanska
\nopagebreak
\begin{center}
\parbox{15.5cm}{\sl\samepage
Department of Physics, University of Lancaster, Lancaster LA1 4YB,
United Kingdom$^{10}$}
\end{center}\end{sloppypar}
\vspace{2mm}
\begin{sloppypar}
\noindent
O.~van~der~Aa,
C.~Delaere,
V.~Lemaitre
\nopagebreak
\begin{center}
\parbox{15.5cm}{\sl\samepage
Institut de Physique Nucl\'eaire, D\'epartement de Physique, Universit\'e Catholique de Louvain, 1348 Louvain-la-Neuve, Belgium}
\end{center}\end{sloppypar}
\vspace{2mm}
\begin{sloppypar}
\noindent
U.~Blumenschein,
F.~H\"olldorfer,
K.~Jakobs,
F.~Kayser,
K.~Kleinknecht,
A.-S.~M\"uller,
G.~Quast,$^{6}$
B.~Renk,
H.-G.~Sander,
S.~Schmeling,
H.~Wachsmuth,
C.~Zeitnitz,
T.~Ziegler
\nopagebreak
\begin{center}
\parbox{15.5cm}{\sl\samepage
Institut f\"ur Physik, Universit\"at Mainz, D-55099 Mainz, Germany$^{16}$}
\end{center}\end{sloppypar}
\vspace{2mm}
\begin{sloppypar}
\noindent
A.~Bonissent,
P.~Coyle,
C.~Curtil,
A.~Ealet,
D.~Fouchez,
P.~Payre,
A.~Tilquin
\nopagebreak
\begin{center}
\parbox{15.5cm}{\sl\samepage
Centre de Physique des Particules de Marseille, Univ M\'editerran\'ee,
IN$^{2}$P$^{3}$-CNRS, F-13288 Marseille, France}
\end{center}\end{sloppypar}
\vspace{2mm}
\begin{sloppypar}
\noindent
F.~Ragusa
\nopagebreak
\begin{center}
\parbox{15.5cm}{\sl\samepage
Dipartimento di Fisica, Universit\`a di Milano e INFN Sezione di
Milano, I-20133 Milano, Italy.}
\end{center}\end{sloppypar}
\vspace{2mm}
\begin{sloppypar}
\noindent
A.~David,
H.~Dietl,
G.~Ganis,$^{27}$
K.~H\"uttmann,
G.~L\"utjens,
W.~M\"anner,
\mbox{H.-G.~Moser},
R.~Settles,
G.~Wolf
\nopagebreak
\begin{center}
\parbox{15.5cm}{\sl\samepage
Max-Planck-Institut f\"ur Physik, Werner-Heisenberg-Institut,
D-80805 M\"unchen, Germany\footnotemark[16]}
\end{center}\end{sloppypar}
\vspace{2mm}
\begin{sloppypar}
\noindent
J.~Boucrot,
O.~Callot,
M.~Davier,
L.~Duflot,
\mbox{J.-F.~Grivaz},
Ph.~Heusse,
A.~Jacholkowska,$^{32}$
C.~Loomis,
L.~Serin,
\mbox{J.-J.~Veillet},
J.-B.~de~Vivie~de~R\'egie,$^{28}$
C.~Yuan
\nopagebreak
\begin{center}
\parbox{15.5cm}{\sl\samepage
Laboratoire de l'Acc\'el\'erateur Lin\'eaire, Universit\'e de Paris-Sud,
IN$^{2}$P$^{3}$-CNRS, F-91898 Orsay Cedex, France}
\end{center}\end{sloppypar}
\vspace{2mm}
\begin{sloppypar}
\noindent
G.~Bagliesi,
T.~Boccali,
L.~Fo\`a,
A.~Giammanco,
A.~Giassi,
F.~Ligabue,
A.~Messineo,
F.~Palla,
G.~Sanguinetti,
A.~Sciab\`a,
R.~Tenchini,$^{1}$
A.~Venturi,$^{1}$
P.G.~Verdini
\samepage
\begin{center}
\parbox{15.5cm}{\sl\samepage
Dipartimento di Fisica dell'Universit\`a, INFN Sezione di Pisa,
e Scuola Normale Superiore, I-56010 Pisa, Italy}
\end{center}\end{sloppypar}
\vspace{2mm}
\begin{sloppypar}
\noindent
O.~Awunor,
G.A.~Blair,
G.~Cowan,
A.~Garcia-Bellido,
M.G.~Green,
L.T.~Jones,
T.~Medcalf,
A.~Misiejuk,
J.A.~Strong,
P.~Teixeira-Dias
\nopagebreak
\begin{center}
\parbox{15.5cm}{\sl\samepage
Department of Physics, Royal Holloway \& Bedford New College,
University of London, Egham, Surrey TW20 OEX, United Kingdom$^{10}$}
\end{center}\end{sloppypar}
\vspace{2mm}
\begin{sloppypar}
\noindent
R.W.~Clifft,
T.R.~Edgecock,
P.R.~Norton,
I.R.~Tomalin
\nopagebreak
\begin{center}
\parbox{15.5cm}{\sl\samepage
Particle Physics Dept., Rutherford Appleton Laboratory,
Chilton, Didcot, Oxon OX11 OQX, United Kingdom$^{10}$}
\end{center}\end{sloppypar}
\vspace{2mm}
\begin{sloppypar}
\noindent
\mbox{B.~Bloch-Devaux},
D.~Boumediene,
P.~Colas,
B.~Fabbro,
E.~Lan\c{c}on,
\mbox{M.-C.~Lemaire},
E.~Locci,
P.~Perez,
J.~Rander,
B.~Tuchming,
B.~Vallage
\nopagebreak
\begin{center}
\parbox{15.5cm}{\sl\samepage
CEA, DAPNIA/Service de Physique des Particules,
CE-Saclay, F-91191 Gif-sur-Yvette Cedex, France$^{17}$}
\end{center}\end{sloppypar}
\vspace{2mm}
\begin{sloppypar}
\noindent
N.~Konstantinidis,
A.M.~Litke,
G.~Taylor
\nopagebreak
\begin{center}
\parbox{15.5cm}{\sl\samepage
Institute for Particle Physics, University of California at
Santa Cruz, Santa Cruz, CA 95064, USA$^{22}$}
\end{center}\end{sloppypar}
\vspace{2mm}
\begin{sloppypar}
\noindent
C.N.~Booth,
S.~Cartwright,
F.~Combley,$^{31}$
P.N.~Hodgson,
M.~Lehto,
L.F.~Thompson
\nopagebreak
\begin{center}
\parbox{15.5cm}{\sl\samepage
Department of Physics, University of Sheffield, Sheffield S3 7RH,
United Kingdom$^{10}$}
\end{center}\end{sloppypar}
\vspace{2mm}
\begin{sloppypar}
\noindent
K.~Affholderbach,$^{23}$
A.~B\"ohrer,
S.~Brandt,
C.~Grupen,
J.~Hess,
A.~Ngac,
G.~Prange,
U.~Sieler
\nopagebreak
\begin{center}
\parbox{15.5cm}{\sl\samepage
Fachbereich Physik, Universit\"at Siegen, D-57068 Siegen, Germany$^{16}$}
\end{center}\end{sloppypar}
\vspace{2mm}
\begin{sloppypar}
\noindent
C.~Borean,
G.~Giannini
\nopagebreak
\begin{center}
\parbox{15.5cm}{\sl\samepage
Dipartimento di Fisica, Universit\`a di Trieste e INFN Sezione di Trieste,
I-34127 Trieste, Italy}
\end{center}\end{sloppypar}
\vspace{2mm}
\begin{sloppypar}
\noindent
H.~He,
J.~Putz,
J.~Rothberg
\nopagebreak
\begin{center}
\parbox{15.5cm}{\sl\samepage
Experimental Elementary Particle Physics, University of Washington, Seattle,
WA 98195 U.S.A.}
\end{center}\end{sloppypar}
\vspace{2mm}
\begin{sloppypar}
\noindent
S.R.~Armstrong,
K.~Berkelman,
K.~Cranmer,
D.P.S.~Ferguson,
Y.~Gao,$^{29}$
S.~Gonz\'{a}lez,
O.J.~Hayes,
H.~Hu,
S.~Jin,
J.~Kile,
P.A.~McNamara III,
J.~Nielsen,
Y.B.~Pan,
\mbox{J.H.~von~Wimmersperg-Toeller}, 
W.~Wiedenmann,
J.~Wu,
Sau~Lan~Wu,
X.~Wu,
G.~Zobernig
\nopagebreak
\begin{center}
\parbox{15.5cm}{\sl\samepage
Department of Physics, University of Wisconsin, Madison, WI 53706,
USA$^{11}$}
\end{center}\end{sloppypar}
\vspace{2mm}
\begin{sloppypar}
\noindent
G.~Dissertori
\nopagebreak
\begin{center}
\parbox{15.5cm}{\sl\samepage
Institute for Particle Physics, ETH H\"onggerberg, 8093 Z\"urich,
Switzerland.}
\end{center}\end{sloppypar}
}
\footnotetext[1]{Also at CERN, 1211 Geneva 23, Switzerland.}
\footnotetext[2]{Now at Fermilab, PO Box 500, MS 352, Batavia, IL 60510, USA}
\footnotetext[3]{Also at Dipartimento di Fisica di Catania and INFN Sezione di
 Catania, 95129 Catania, Italy.}
\footnotetext[4]{Now at LBNL, Berkeley, CA 94720, U.S.A.}
\footnotetext[5]{Also Istituto di Cosmo-Geofisica del C.N.R., Torino,
Italy.}
\footnotetext[6]{Now at Institut f\"ur Experimentelle Kernphysik, Universit\"at Karlsruhe, 76128 Karlsruhe, Germany.}
\footnotetext[7]{Supported by CICYT, Spain.}
\footnotetext[8]{Supported by the National Science Foundation of China.}
\footnotetext[9]{Supported by the Danish Natural Science Research Council.}
\footnotetext[10]{Supported by the UK Particle Physics and Astronomy Research
Council.}
\footnotetext[11]{Supported by the US Department of Energy, grant
DE-FG0295-ER40896.}
\footnotetext[12]{Now at Departement de Physique Corpusculaire, Universit\'e de
Gen\`eve, 1211 Gen\`eve 4, Switzerland.}
\footnotetext[13]{Supported by the Commission of the European Communities,
contract ERBFMBICT982874.}
\footnotetext[14]{Supported by the Leverhulme Trust.}
\footnotetext[15]{Permanent address: Universitat de Barcelona, 08208 Barcelona,
Spain.}
\footnotetext[16]{Supported by Bundesministerium f\"ur Bildung
und Forschung, Germany.}
\footnotetext[17]{Supported by the Direction des Sciences de la
Mati\`ere, C.E.A.}
\footnotetext[18]{Supported by the Austrian Ministry for Science and Transport.}
\footnotetext[19]{Now at SAP AG, 69185 Walldorf, Germany}
\footnotetext[20]{Now at Groupe d' Astroparticules de Montpellier, Universit\'e de Montpellier II, 34095 Montpellier, France.}
\footnotetext[21]{Now at BNP Paribas, 60325 Frankfurt am Mainz, Germany}
\footnotetext[22]{Supported by the US Department of Energy,
grant DE-FG03-92ER40689.}
\footnotetext[23]{Now at Skyguide, Swissair Navigation Services, Geneva, Switzerland.}
\footnotetext[24]{Also at Dipartimento di Fisica e Tecnologie Relative, Universit\`a di Palermo, Palermo, Italy.}
\footnotetext[25]{Now at McKinsey and Compagny, Avenue Louis Casal 18, 1203 Geneva, Switzerland.}
\footnotetext[26]{Now at Honeywell, Phoenix AZ, U.S.A.}
\footnotetext[27]{Now at INFN Sezione di Roma II, Dipartimento di Fisica, Universit\`a di Roma Tor Vergata, 00133 Roma, Italy.}
\footnotetext[28]{Now at Centre de Physique des Particules de Marseille, Univ M\'editerran\'ee, F-13288 Marseille, France.}
\footnotetext[29]{Also at Department of Physics, Tsinghua University, Beijing, The People's Republic of China.}
\footnotetext[30]{Now at SLAC, Stanford, CA 94309, U.S.A.}
\footnotetext[31]{Deceased.}
\footnotetext[32]{Also at Groupe d' Astroparticules de Montpellier, Universit\'e de Montpellier II, 34095 Montpellier, France.}  
\footnotetext[33]{Now at University of Florida, Department of Physics, Gainesville, Florida 32611-8440, USA}
\footnotetext[34]{Now at Institut Inter-universitaire des hautes Energies (IIHE), CP 230, Universit\'{e} Libre de Bruxelles, 1050 Bruxelles, Belgique}
\setlength{\parskip}{\saveparskip}
\setlength{\textheight}{\savetextheight}
\setlength{\topmargin}{\savetopmargin}
\setlength{\textwidth}{\savetextwidth}
\setlength{\oddsidemargin}{\saveoddsidemargin}
\setlength{\topsep}{\savetopsep}
\normalsize
\newpage
\pagestyle{plain}
\setcounter{page}{1}


\cleardoublepage 
\newpage 
\setlength{\topmargin}{-1cm}
\setlength{\oddsidemargin}{-0.5cm}
\newpage
\pagestyle{plain}
\setcounter{page}{1}
%
%
\pagenumbering{arabic}
\normalsize
\setlength{\textheight}{23cm}
\setlength{\textwidth}{18cm}

\setlength{\topmargin}{-1cm}
\setlength{\oddsidemargin}{-0.5cm}

\section{Introduction}
\label{intro} 
Unlike at LEP\,1 energies~\cite{Hlep1}, ALEPH searches for the Standard
       Model Higgs boson at LEP\,2~\cite{Hlep2001} were performed under the
       assumption that the  Higgs boson decays predominantly into ${\rm b\bar b}$. 
       Invisible final states, which would arise
       for instance from a decay
       into a neutralino pair, were also investigated~\cite{Hlep2001}.
       It is also possible, in the MSSM~\cite{carena} as well as
       in more general two-Higgs-doublet models,
       to find parameter sets for
       which the decay into ${\rm b\bar b}$ is strongly suppressed, to the
       benefit of other decay modes such as ${\rm c\bar c}$, gg or
       $\tau^+\tau^-$.
       It is shown in this letter that either existing or
       slightly modified ALEPH searches for the Higgsstrahlung process
       ${\rm e}^+{\rm e}^- \to {\rm HZ}$ are sensitive to these decays.
       These analyses are based on the $630\,{\rm pb}^{-1}$ of
       data collected by ALEPH between 1998 and 2000 at centre-of-mass
       energies ranging from 189 to 209\,GeV (Table~\ref{tab_lumi}).

      \begin{table}[h]
      \begin{center}
      \caption{\footnotesize Integrated luminosities, centre-of-mass energy
    ranges
      and mean centre-of-mass energy values for data collected by the ALEPH
      detector from 1998  to 2000.}
      \begin{tabular}{|c|c|c|c|}
        \multicolumn{4}{c}{} \\  \hline \hline
       Year & Luminosity (${\rm pb}^{-1}$) & Energy range (GeV)
            & $\langle \sqrt{s} \rangle$ (GeV) \\ \hline
       2000 &   11.2            & $207-209$        & 208.0 \\
            &  122.6            & $206-207$        & 206.6 \\
            &   80.0            & $204-206$        & 205.2 \\ \hline
       1999 &   45.2            &   $-$              & 201.6 \\
            &   86.3            &   $-$              & 199.5 \\
            &   79.9            &   $-$              & 195.5 \\
            &   28.9            &   $-$              & 191.6 \\ \hline
       1998 &  176.2            &   $-$              & 188.6 \\
      \hline
      \hline
      \end{tabular}
      \label{tab_lumi}
      \end{center}
      \end{table}

       This letter is organized as follows. A brief description of the
       ALEPH detector is given in Section~2. The event selections pertaining
       to a flavour-independent search for the Higgs boson produced via the
       Higgsstrahlung process are examined in turn in Section~3, and the
       results are summarized in Section~4.

\section{ALEPH Detector}
 A detailed description of the ALEPH detector and its performance can be found 
in Refs.~\cite{det1} and~\cite{det2}. The tracking system consists of a silicon vertex detector,
 a cylindrical drift chamber and a large time projection chamber (TPC), immersed
 in a 1.5\,T axial magnetic
 field provided by a superconducting solenoidal coil.
With these detectors, 
  a transverse momentum resolution of 
$\delta p_{\rm t}/p_{\rm t} =6 \times 10^{-4} p_{\rm t} \oplus 5 \times 10^{-3}$ ($ p_{\rm t}$ in GeV/$c$) is
achieved.

 An electromagnetic calorimeter placed between the TPC and the superconducting coil identifies
 electrons and photons, and measures their energies with a resolution of
  \mbox{$\delta E/E=0.18/\sqrt{E} +0.009$ ($E$ in GeV)}.
The iron return yoke is instrumented with 23 layers of streamer tubes and serves
 as a hadron calorimeter and muon filter.
 Two additional  double layers of streamer 
tubes outside the return yoke aid the identification of muons.

An energy flow algorithm~\cite{det2} combines the information from the tracking detectors and 
the calorimeters and provides a list of reconstructed charged and neutral particles.
The  achieved energy resolution  is  $\sigma(E)=0.6 \sqrt{E} +0.6$ ($E$ in\,GeV).

\section{Event selections} 
\label{analysis} 
    In this analysis, decays of the Higgs boson to hadrons or to tau
    pairs are considered. Hadronic Higgs boson decays are 
    searched for in four-jet, missing-energy and leptonic (electron
    or muon pair) final states as for the \SM\ search for the
    ${\rm e}^+{\rm e}^- \to {\rm HZ}$ process. 
    The event selections used here for the missing
    energy and leptonic final states are based on those used in the
    previous searches~\cite{SM_higgs,sm189},
    however without b-tagging information.
    A specially designed flavour-independent
    selection is used for the four-jet channel. Higgs
    boson decays to a tau pair are searched for in the final state
    $\tau^+\tau^- \rm {q \bar{q}}$, using the same selection 
    described in Ref.~\cite{sm189}.

    Signal efficiencies and background contributions from \SM\
    processes are estimated with simulated event samples which include
    a full simulation of the ALEPH detector.  To study the 
    signal efficiency for \Hhad{},
    events from the HZ process are generated in which
    the H decays to \bbbar,  \ccbar\ or \gg\  and the Z  into a pair
    of quarks, neutrinos, electrons or muons.  Events in which
    H decays to a pair of taus and the Z  to a pair of quarks
    are used to determine the tau  channel efficiency. Signal events are
    simulated with the Monte Carlo generator \HZHA~\cite{HZHA},
    for \mH\ from 40 to 115\mc2 in steps of 5\mc2.
    The simulated background
    event samples are identical to those used in Ref.~\cite{SM_higgs}.
 
\subsection{Leptonic and missing energy final states}
    The leptonic channel event selection, which does
    not include b-tagging information, 
    is unchanged with respect to Ref.~\cite{SM_higgs}.
    The reconstructed
    Higgs boson mass, computed  as the mass recoiling against 
    the lepton system, is used
    as a discriminant variable in the confidence level calculation.
    When this analysis is applied to the data~\cite{Hlep2001,sm189,zz},
    70 events are observed,
    in agreement with the 73.4 events expected from the \SM\
    backgrounds.  The signal efficiencies for 
    ${\rm H} \to {\rm  b\bar{b}}$, \ccbar\ or \gg\
    are found to be quite similar, at about 80\%
    over the entire mass range, 
    except when approaching
    the kinematic limit for HZ production where it falls to  40\%.

    As for the leptonic analysis, the \SM\ missing energy
    event selection applied to data collected in 2000 does not include
    b-tagging information, and is therefore used in this search.
    In this analysis, the reconstructed Higgs boson mass is used as a 
    discriminant variable in the confidence level calculation.
    Prior
    to 2000, the \SM\ missing energy event selections relied
    explicitly on b tagging, and are therefore inappropriate for this
    analysis.  A modified version of the three-neural-network analysis
    described in Ref.~\cite{sm189} is applied to this data sample.  This
    analysis uses the seven-variable anti-$\rm q\bar{q}$ and three-variable 
    anti-WW neural networks
    used in the standard analysis.
    When these analyses are applied to the data, 177 events
    are selected in the sample  with 181 expected
    from \SM\ background processes.     Cuts on the two neural network outputs are chosen to optimize the
    search sensitivity as a function of 
    the  Higgs boson mass.
    The signal efficiencies for  $\rm H \to {\rm  b\bar{b}}$, \ccbar\ or \gg\ are found to
    be quite similar, and are about 40\% over the entire
    mass range, falling to 20\% near the kinematic limit.

\subsection{Final state with taus}
The search for the $\tau^+ \tau^- \rm {q \bar{q}}$
    final state of Ref.~\cite{SM_higgs} is used here 
    for Higgs bosons decaying into tau
    pairs.
    The selection
    efficiency  is around 40\%. 
    The reconstructed Higgs boson mass is used as a discriminant
    variable in the confidence level calculation.  A total of 27 candidate
    events is selected in the data, in agreement with 27.2 expected
    from background processes.

\subsection{Four-jet final state} 
    The preselection of the flavour-independent four-jet selection is similar 
    to that of the standard four-jet analysis~\cite{zz}.
    Events are required to have at least
    eight charged particle tracks,        
    and the total energy of the charged particles must be larger than
    10\% of the centre-of-mass energy.
    Events from
    radiative returns to the Z resonance, in which a photon escapes
    undetected down the
    beam pipe, are rejected by requiring the momentum $p_z$ 
    of the event along the
    beam axis to satisfy $p_z < 1.5(m_{\rm {vis}}-90)$, where $m_{\text{vis}}$ is
    the total visible mass in  the event, expressed in~\mc2.  
    Events are then clustered into
    four jets using the Durham jet-clustering algorithm~\cite{durham}. 
    The transition from four to three jets is required to  occur for 
    $y_{34} > 0.008$.
    Events from radiative returns to the Z with
    a photon in the detector are rejected if
    more than 80\% of the energy of any jet is in the form of electrons and
    photons.  Events from
    semileptonic decays of \WW{} are rejected by requiring that the energy
    of the most energetic identified electron or muon is less than 20\,GeV.
    To avoid overlap with the leptonic selection, events containing a pair
    of identified electrons or muons with an invariant mass greater than
    40\mc2\ are rejected.
    After this preselection, signal efficiencies for
     ${\rm H} \to {\rm b \bar{b}}$,  ${\rm c \bar{c}}$
    and
    $ {\rm gg}$ 
     are of the order of 70\%.
    The numbers of events expected from  background processes
    and the numbers of candidate events observed in the data
    are reported in Table~\ref{tab_bkg}.  The comparison indicates a reasonable
    agreement between data and the expectation from \SM\ processes
    at the preselection level.

      \begin{table}[h]
      \begin{center}
      \caption[]{\footnotesize Numbers of expected events from
 background processes and numbers of candidate events  collected at centre-of-mass
 energies from 189 to 209\,GeV, at the preselection level  for the four-jet 
 final state selection.\rm}
      \begin{tabular}{|c|c|c|c|c|c|} 
      \multicolumn{6}{c}{} \\  \hline \hline
 $\sqrt{s}$ (GeV) & \multicolumn{4}{c|}{Background process contributions}     & Data \cr\cline{2-5}
   & WW & $\rm q\bar{q}$ & ZZ &  Total  & \\
 \hline
 188.6 & 1002.1 & 261.5 & 63.8& 1327.5  & 1242 \\
 \hline
 191.6 & 165.2& 41.9& 11.8& 218.9 & 221 \\
 \hline
 195.5 & 459.8& 108.1& 35.7& 603.4 & 614 \\
 \hline
 199.5 & 492.3& 108.0& 40.0& 640.1 & 624 \\
 \hline
 201.6 & 238.9& 51.2& 19.7& 309.7 & 261  \\
 \hline
 204--209 & 1251.0& 247.9 & 102.8& 1601.6 & 1601 \\
 \hline
 All $\sqrt{s}$ & 3609.3& 818.6 &273.8 & 4701.7 &4563 \\
 \hline \hline
 \end{tabular}
\label{tab_bkg}
\end{center}
 \end{table}

    This analysis uses several of the
    kinematic variables used in the standard analysis and in addition,  three
    variables based on di-jet mass information:

\begin{itemize} 
\item[$\bullet$]  the significance of the distance to the \WW{} hypothesis, defined as 
$$\mbox{\EllWi~=~\minijkl\,\{(\mij +\mkl $-2$\mw)$^2$/$\sigma_s^2$ + 
(\mij $-$\mkl)$^2$/$\sigma_d^2$}\},$$ 
where $\sigma_s$~=~4\mc2\ and $\sigma_d$~=~10\mc2\  are the resolutions on the sum and the
difference of the  di-jet masses for \WpWm\ production, and $i,j,k,l$ denote the 
four jets reconstructed in the event;
\item[$\bullet$]   the probability density functions \S_dens(\EllhZi) and \B_dens(\EllhZi) for signal and background. Here, \EllhZi\ is the significance of the distance to the HZ hypothesis. It depends on \mH\ and  is defined as

   $$\mbox{\EllhZi~=\minijkl\,$\left\{\displaystyle\frac{\left[(m_{ij}+m_{kl})-(m_{\rm Z}+m_{\rm H})
\right]^2}{\sigma^2_{\Sigma}} +
                      \frac{\left[(m_{ij}-m_{kl})-(m_{\rm H}-m_{\rm Z})\right]^2}{\sigma^2_{\Delta}}\right\}$},$$

  where $\sigma_\Sigma$ and $\sigma_\Delta$ are the resolutions on the sum
    and the difference of di-jet masses for HZ production.
    Simulated event
    samples are used to parametrize \S_dens and \B_dens.

\end{itemize} 
    These three variables are combined with the smallest jet energy \Ejmin{}, 
     the largest jet energy \Ejmax{}, and the product \Ejminthe{}
    of the smallest jet energy and the minimum angle between
    any two jets, in a six-variable neural network.  In order to
    optimize the performance over a wide range of masses, separate neural networks
    are trained for several Higgs boson mass hypotheses, ranging from
    40 to 115\mc2, in steps of 5\mc2, at three different
    centre-of-mass energies~: 189, 199.5 and 206.7\,GeV.  For each of these
    networks, an optimization using the \N95~\cite{nbar95} prescription
    is performed to determine the appropriate cut for the neural network output.
    As the di-jet mass information is included in the
    neural network, it is not included again as a discriminant variable
    when computing the confidence level.

    As neural networks are trained every 5\mc2,
    a sliding
    method is used to determine the selection at intermediate Higgs boson masses.
    For a mass $m_{\rm H}$ intermediate between two training masses
    $m_1$ and $m_2$, ${\cal S}_{m_H}$ (${\cal B}_{m_H}$) is interpolated from ${\cal S}_{m_1}$
    (${\cal B}_{m_1}$) and
    ${\cal S}_{m_2}$ (${\cal B}_{m_2}$).
    These quantities are input to the
    network trained at $m_1$ and to the network trained at $m_2$. 
    The two neural networks outputs are then interpolated to calculate $NN_{m_{\tilde{\rm H}}}$.
    The cut value on $NN_{m_{\tilde{\rm H}}}$, the 
    signal efficiency and
    background expectation at the intermediate mass are similarly obtained by interpolations.
    The validity of this procedure has been established
    with test samples of signal events simulated at $\sqrt{s}$ = 199.5\,GeV 
    for masses  between 60 and 100\mc2\ in steps of~1~\mc2.

\begin{figure}[tb]
\begin{center}
\begin{picture}(100,80)
\put(20,70){ALEPH } 
\psfrag{b1}[t][b][.9]{~\hspace{-23mm}$m_{\rm H}$ (\mc2)} 
\psfrag{b2}[b][t][1]{~\hspace{-30mm} number of events}
\put(0,-5){\epsfxsize90mm\epsfbox{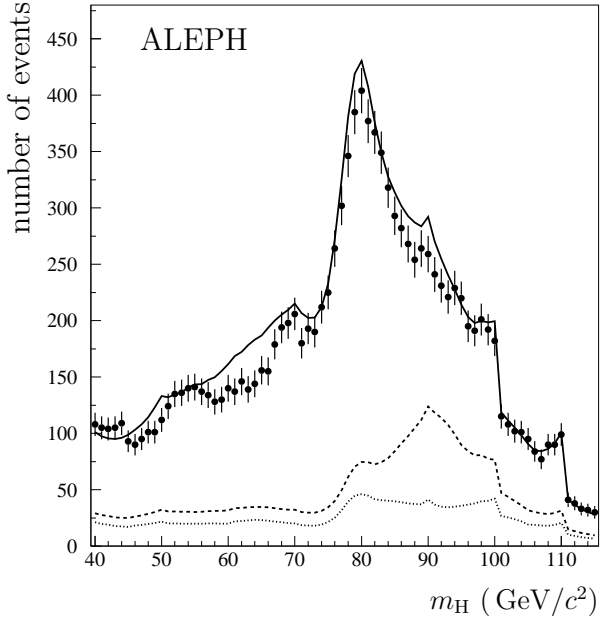}}
\end{picture}
\vspace{0.5cm}
\caption{\footnotesize Expected background (solid curve) compared to 
collected data (dots) by the four-jet 
selection as a  function of the Higgs boson mass hypothesis
at centre-of-mass energies from 189 to 209\,GeV. The dotted curve represents the 
contribution from the $\rm q\bar{q}$ background and the dashed curve 
represents the additionnal contribution from the ZZ background.
\label{4jets_bdfmh_indflavor}} 
\end{center}
\end{figure}

    The number of events selected by this analysis in data is compared to
    the background expectation  in Fig.~\ref{4jets_bdfmh_indflavor} as
    a function of the Higgs boson mass hypothesis.  
    The
    points are statistically correlated, as mass resolution
    effects usually allow the events to contribute to several adjacent mass bins.
    As a result, a deficit in one mass region can be propagated to a
    large range of mass hypotheses, as observed in the 60 and 90\mc2\ regions,
    correlated to the~\WW{} and ZZ deficits already described in Ref.~\cite{zz}
    and Ref.~\cite{ww}.

\section{Results}
In every channel under investigation,
no departure from  \SM\ expectations consistent with
 the presence of a Higgs signal is
observed in the data. Lower limits on the lightest scalar 
Higgs boson mass are
derived as a function of  \xihad\ 
or  \xitau, 
 the  product of the branching fraction to either hadronic jets or tau pairs
    and of the ratio of the production cross section to the \SM\
    production cross section.
   In order to obtain a flavour-independent limit for the
   decay  \Hhad{}, the smallest of the signal efficiencies for 
   $\rm {H} \to $ \bbbar
   ,\, \ccbar\ and \gg\
   is used at each Higgs boson mass hypothesis.
   For a Higgsstrahlung cross section equal to that of the \SM\
   and for 100\% branching fraction to hadrons,
   Higgs boson masses below  110.6\mc2\ are excluded
   at 95\% C.L., where a limit of 110.5\mc2\  is expected in the
   absence of signal.

   When the  parameter \xihad\ is allowed to
   vary, the result of the flavour-independent
   search is expressed as an excluded domain in the \xihadplane\
   plane, as shown in Fig.~2a.
  Results from
   Ref.~\cite{lep1} are used to exclude Higgs boson mass hypotheses below
   40\mc2.

    A similar procedure is followed to obtain a limit on the decay
    $\rm{H}  \to \tau^+\tau^-$, and
     the exclusion in the \xitauplane\ plane is shown
    in Fig.~2b.
    For \xitau = 1, a lower limit on the Higgs boson mass of
    112.4\mc2\ at 95\% C.L. is obtained, where a limit of
    113.9\mc2\ is expected in the absence of signal.
    Under the assumption that  \xihad\ $+$ \xitau $= 1$, a 
    109.1\mc2 lower limit on \mH\ is obtained irrespective of \xitau.

    The dominant systematic error sources, evaluated as described in
    Ref.~\cite{sm189}, are included in the obtained limits.
    The finite size of the
    simulated event samples, the jet energy and angular resolutions and
    the uncertainties in the signal and background cross section
    estimations affect all  the topologies under investigation.  In
    the leptonic channel, lepton identification and isolation are additional
    sources of uncertainty.  For the four-jet channel, systematic
    uncertainties due to differences between data and simulation in the
    event selection variables are taken into account with an event
    reweighting method~\cite{weight}.
    The global effect
    of these uncertainties is to decrease the hadronic
    limit by 190\mevc2, and the tau limit by 10\mevc2.

\section{Conclusions}
  In order to explore nonstandard Higgs scenarios, searches for Higgs
    bosons produced via Higgsstrahlung decaying to hadrons and to tau leptons were
    performed. The selections are similar to those used in previous searches,
    except for the search in the four-jet final state, where a new analysis
    was designed in order to cope with hadronic Higgs boson decays without a flavour
    tag.  No evidence of Higgs boson production is observed in the search for
    either 
    hadronic or tau decays in the data collected at energies
    between 189 and 209\,GeV. For a \SM\ Higgsstrahlung cross section and
    a 100\% branching fraction to hadrons, masses below 110.6\mc2\ are
    excluded at 95\% C.L. independent of the flavour of the Higgs boson decay.
    Results on flavour-independent Higgs boson production have also been   
    reported by the OPAL collaboration~\cite{others} with lower energy data. For a
    \SM\ Higgsstrahlung cross section and a 100\% branching fraction
    to $\tau^+\tau^-$, masses below 112.4\mc2\ are excluded at~95\%~C.L..

\section*{Acknowledgements}
We congratulate our colleagues from the accelerator divisions for the very successful running
of LEP at high energies. We  are indebted to the engineers and technicians in all our institutions for the excellent performance of ALEPH.
 Those of us from non-member countries thank CERN for its hospitality.
\begin{figure}[tb]
\begin{center}
\begin{picture}(160,80)
\psfrag{a1}[t][b][.9]{~\hspace{-23mm}$m_{\rm H}$ (\mc2)} 
\psfrag{a2}[b][t][1]{~\hspace{-0mm} \xitau}
\psfrag{a3}[t][b][0.9]{~\hspace{-23mm}$m_{\rm H}$ (\mc2)} 
\psfrag{a4}[b][t][1]{~\hspace{-3mm} \xihad}

\put(-10,-10){\epsfxsize90mm\epsfbox{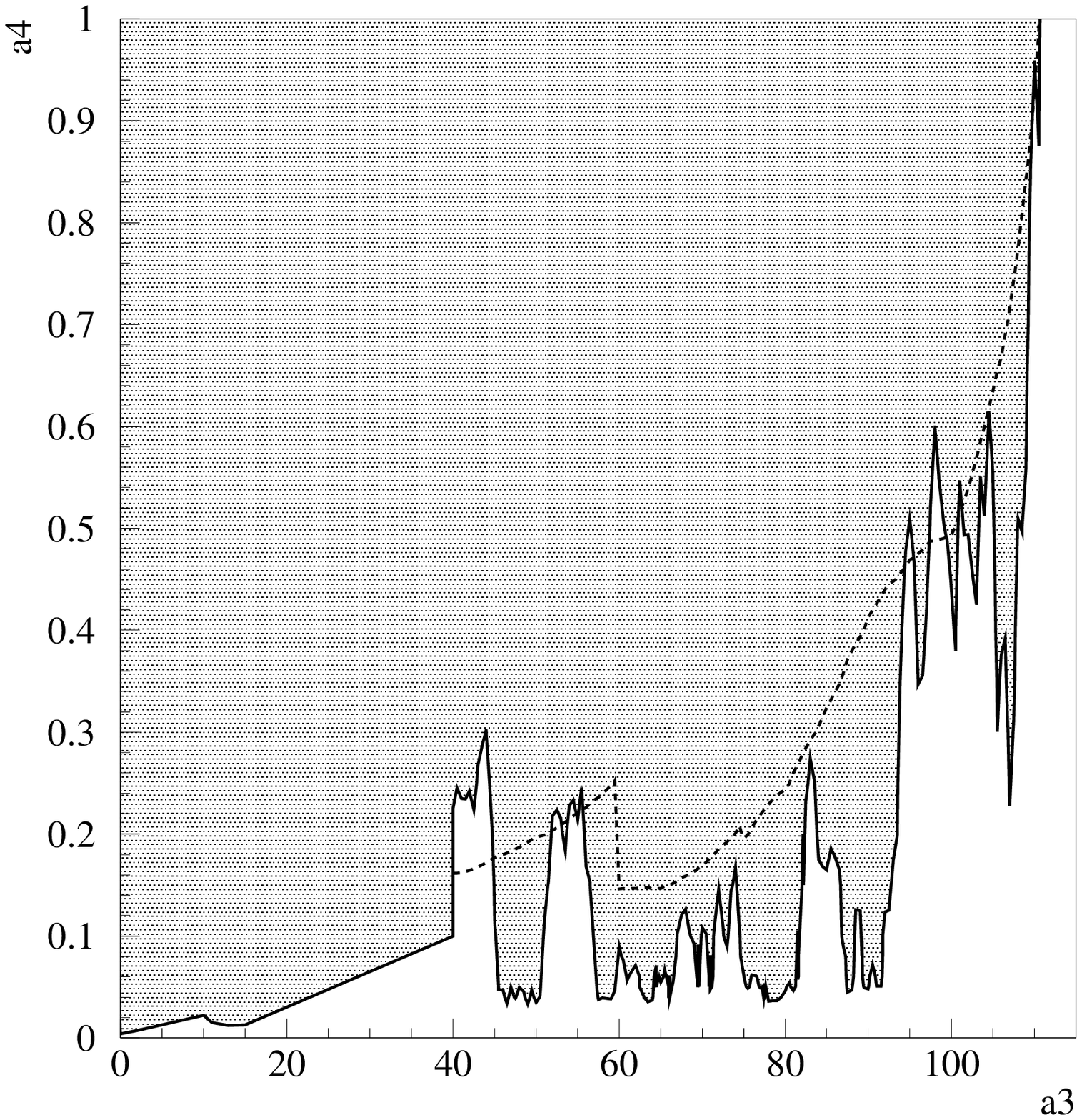}}
\put(80,-10){\epsfxsize90mm\epsfbox{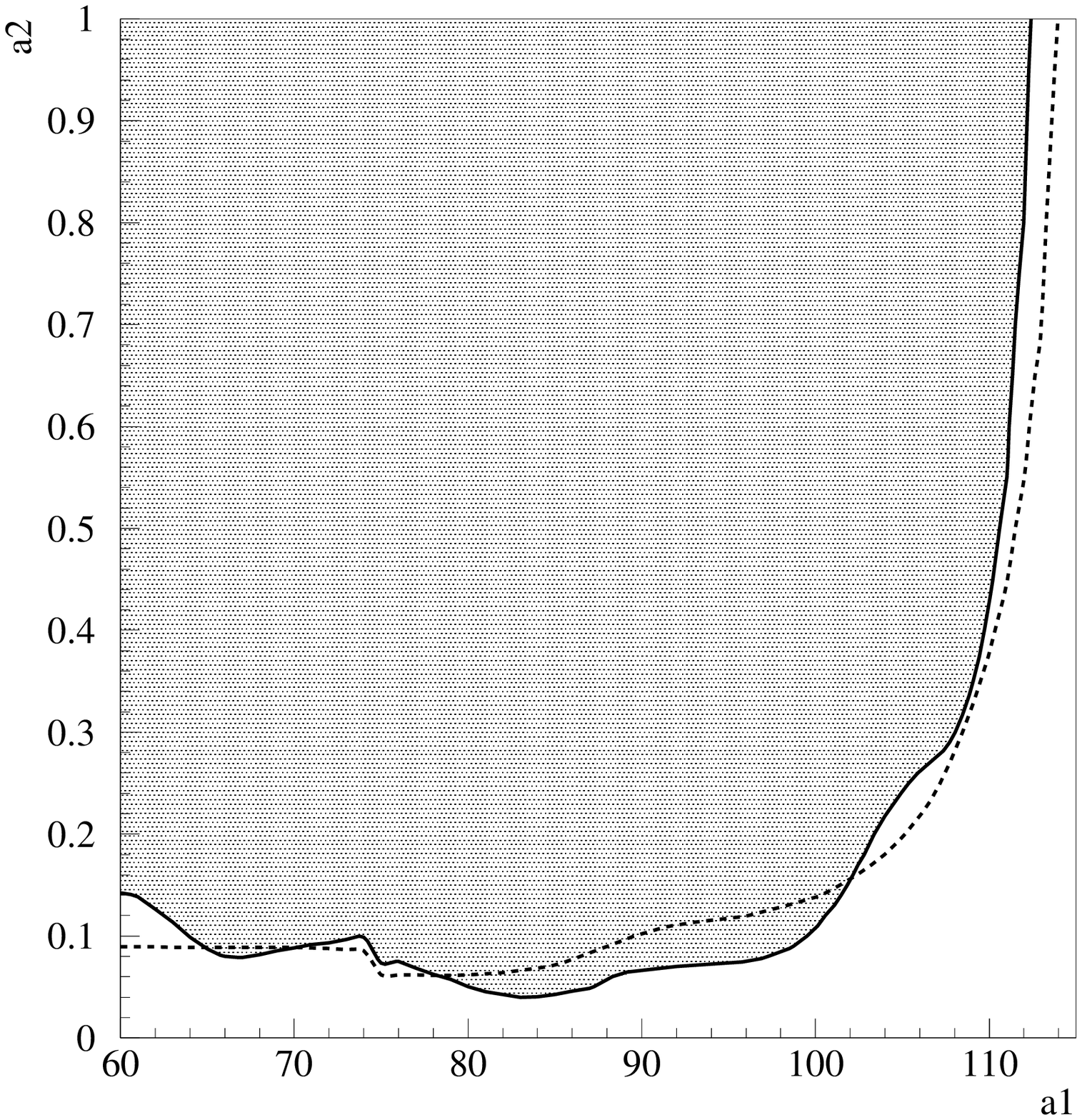}}
\put(65,2){(a)}
\put(155,2){(b)}
\put(5,65){ALEPH } 
\put(95,65){ALEPH } 
\end{picture}\end{center}
\vspace{0.5cm}
\caption{\footnotesize
 Expected (dashed line) and observed (shaded area)  95\% C.L.
limits on a) \xihad\  and b) \xitau,
as a function of the Higgs boson mass hypothesis and for 
centre-of-mass energies up to 209\,GeV.
\label{cl_all}} 
\end{figure}

\end{document}